\pdfoutput=1
\documentclass[letter]{article}

\usepackage{latexsym,epsfig,graphicx,amsmath,amssymb,
	amscd,undertilde,multirow,chicago,psfrag,paralist,dsfont,url}
\usepackage[titletoc]{appendix}
\usepackage{booktabs}
\usepackage{lscape}
\usepackage{footnote}
\usepackage{authblk}
\graphicspath{{figures/}}
\newcommand{\thetavec}{{\boldsymbol{\theta}}}

\newcommand{\betavec}{{\boldsymbol{\beta}}}
\newcommand{\gammavec}{{\boldsymbol{\gamma}}}
\newcommand{\etavec}{{\boldsymbol{\eta}}}

\newcommand{\E}{\mathbf{E}}

\newcommand{\Hset}{\mathbf{H}}

\newcommand{\pr}{{\rm Pr}}

\newcommand{\NOR}{{\rm N}}

\newcommand{\wh}{\widehat}

\newcommand{\Xvec}{\boldsymbol{X}}
\newcommand{\Zvec}{\boldsymbol{Z}}

\newcommand{\Iset}{{\boldsymbol{I}}}

\newcommand{\Lvec}{{\boldsymbol{L}}}
\newcommand{\Uvec}{{\boldsymbol{U}}}
\newcommand{\nuvec}{{\boldsymbol{\nu}}}

\newcommand{\muvec}{{\boldsymbol{\mu}}}

\newcommand{\Yvec}{\mathbf{Y}}

\newcommand{\vmat}{\mathbf{V}}

\newcommand{\Dvec}{\mathbf{D}}

\newcommand{\Amat}{\mathbf{A}}

\newcommand{\omegamat}{\mathbf{\Omega}}

\newcommand\given[1][]{\:#1\vert\:}
\def\T{{ \mathrm{\scriptscriptstyle T} }}

\begin{document}
	
	%%%%%%%%%%%%TITLE%%%%%%%%%%%%%%%%%%%%%%%%%%%%%%%%%%%
	\title{Hypothesis Testing for Detecting Outlier Evaluators}
	
	%%%%%%%%%%%%%%%%%%%%%%%%%%%%%%%%%%%%%%%%%%%%%%%%%%%%%%%%%%%%%%%%%%%%%%%%%%%%%%%%%%%%%%%%%%%%%%%%%%%%%%%%%%%%%%%%%
	
	\author[1]{Li Xu}
	\author[1,2,3]{Molin Wang\thanks{Corresponding author: e-mail: stmow@channing.harvard.edu}}
	\affil[1]{Department of Epidemiology, Harvard T.H. Chan School of Public Health, Boston, Massachusetts 02115, USA}
	\affil[2]{Department of Biostatistics, Harvard T.H. Chan School of Public Health, Boston, Massachusetts 02115, USA}
	\affil[3]{Channing Division of Network Medicine, Harvard Medical School and Brigham and Women's Hospital, Boston, Massachusetts 02115, USA}

	%\date{\today}
	\date{}
	\maketitle
	%%%%%%%%%%%%%%%%%%%%%%%%%%%%%%%%%%%%%%%%%%%%%%%%%%%%%%%%%%%%%%%%%%%%%%%%%%%%%%%%%%%%%%%%%%%%%%%%%%%%%%%%%%%%%%%%
	\begin{abstract}
		In epidemiological studies, very often, evaluators obtain measurements of disease outcomes for study participants. In this paper, we propose a two-stage procedure for detecting outlier evaluators. In the first stage, a regression model is fitted to obtain the evaluators' effects. The outlier evaluators are considered as those with different effects compared with the normal evaluators. In the second stage, stepwise hypothesis testings are performed to detect outlier evaluators. The true positive rate and true negative rate of the proposed procedure are assessed in a simulation study. We apply the proposed method to detect potential outlier audiologists among the audiologists who measured hearing threshold levels of the participants in the Audiology Assessment Arm of the Conservation of Hearing Study, which is an epidemiological study for examining risk factors of hearing loss.
	\end{abstract}

	%\tableofcontents
	%\newpage
	\section{Introduction}
	To investigate the relationship between risk factors and disease outcomes, many medical and epidemiological studies rely on evaluators to measure the exposures or outcomes of interest among study participants. Especially in large epidemiological studies, different evaluators are involved in providing measurements \shortcite{sanders2015qualitative,dogan2015clinical,miller2016integrating}. The difference between evaluators has been recognized as a source of measurement errors \shortcite{beckler2018reliability}, and methods to detect the outliers among evaluators are necessary for the data quality control. Many outlier detection techniques have been developed specifically to
	certain application domains. \shortciteN{7972424} use the K-nearest neighborhood-based method the detect fraud in credit card transactions. \shortciteN{dey2022outlier} design an influence measure under a hierarchical generalized linear model for the outliers in neuroimaging data. \shortciteN{huang2021outlier} propose a robust kernel dictionary learning method to detect the outliers in the process-monitoring data from the industrial cyber-physical system. For the outliers among correlated data, \shortciteN{zhu2022gaussian} use an LSTM-based variational autoencoder to detect the anomaly in time series. For the multivariate data, \shortciteN{cabana2021multivariate} propose a Mahalanobis-distance-based detection method using robust shrinkage mean and variance-covariance detection. However, these outlier detection methods are for detecting outliers among observations. They do not apply for detecting outliers among evaluators, which are complicated  by the fact that evaluatees' characteristics may also be key contributors to evaluation results.
	
	Our motivating example in this paper is from the Conservation of Hearing Study (CHEARS) \shortcite{curhan2021osteoporosis}. CHEARS investigates hearing loss risk factors among participants in the Nurses’ Health Study II (NHS II) \shortcite{bao2016origin}, which consists of 116,430 women aged from 25 to 42 at enrollment in 1989. The CHEARS Audiology Assessment Arm (AAA) assessed the longitudinal change in the pure-tone air and bone conduction audiometric hearing thresholds. The threshold was measured in decibels in hearing level (dB HL), across the full range of conventional frequencies (0.5-8 kHz) \shortcite{curhan2020prospective}. At the baseline of AAA, 3,136 participants from the NHS II were completely tested by 46 audiologists. Outlier audiologists tend to give significantly higher or lower hearing measurements; thus, it is necessary to detect potential outlier audiologists in the data collection stage of a study. Detecting outliers makes it possible to implement the early-stage intervention, and consequently provides more reliable data for later data analysis.
	
	In this paper, we develop a two-stage method for detecting outlier evaluators. In the first stage, a multivariate regression model is fitted to estimate the effect of evaluators on the outcome measurements; outlier evaluators are those with different effect estimates from the other evaluators. In the second stage, by extending the many-outlier procedure in \citeN{rosner1975detection} and \citeN{rosner1983percentage}, which was developed to detect outliers among independent observations, outliers are detected based on evaluators' estimated effects from the first stage analysis.
	
	The rest of the paper is organized as follows. In Section 2, we present the two-stage procedure for detecting outliers among evaluators. In Section 3, we present the results of a simulation study for assessing the performance of the proposed method. In Section 4, we illustrate the proposed method by detecting potential outlier audiologists in the baseline measurements of the CHEARS AAA. We end with a discussion in Section 5.
	
	\section{Methods}
	Our outlier detection procedure contains two stages. In the first stage, we fit a linear regression model to estimate the effects of evaluators. In the second stage, we propose a stepwise hypothesis testing procedure that detects one potential outlier at each step.
	
	\subsection{Stage I: Linear Regression Analysis}
	Firstly, we consider the simple scenario in which each participant has only one measurement. Let $i\in\{1,2,\ldots,N\}$ be the index of study participants and $j\in\{1,2,\ldots,M\}$ be the index of evaluators. Let $N_j$ be the number of participants assigned to the $j$-th evaluator and $\sum_{j=1}^{M}N_j=N$. We can assume the following linear regression model:
	\begin{align*}
		Y_i = \sum_{j=1}^{M}\beta_jT_i^{(j)} + \gammavec^{\T} \Xvec_i + \epsilon_i,
	\end{align*}
	where $Y_i$ is the single measurement of the $i$-th participant, $T_i^{(j)}$ is the evaluator indicator such that $T_i^{(j)}=1$ when the $i$-th participant is measured by the $j$-th evaluator and $T_i^{(j)}=0$ otherwise, $\Xvec_i$ is a $q$-dimensional vector containing potential predictors of the outcome, $\gammavec$ is a $q$-dimensional coefficient vector, and $\gammavec^\T$ represents the transpose of $\gammavec$. Without further specification, all vectors are column vectors in this paper. The coefficients $\betavec=(\beta_1,\beta_2,\ldots,\beta_M)^\T$ represent a combination of the intercept and the effects of the $M$ evaluators; for presentational convenience, we will refer $\beta$ as the evaluator effects hereafter. If all $\beta_j$'s are the same for $j=1,\ldots,M$ in the linear model,  the measurement $Y_i$ would be the same no matter which evaluator measures the outcome for the $i$-th participant. %The random noise $\epsilonvec=(\epsilon_1,\ldots,\epsilon_N)^\T$ follows a multivariate normal distribution $\NOR(\zerovec_N,\sigma^2\Imat_N)$, where $\zerovec_N$ is an $N$-element vector of zeros and $\Imat_N$ is $N\times N$ identity matrix. %We further define the joint predictor matrix $\wmat=(\tmat,\xmat)$, where $\tmat=(\tmat_1,\ldots,\tmat_N)^{\T}$ and $\Xvec=(\Xvec_1,\ldots,\Xvec_N)^{\T}$.
	
	In practice, one participant may have multiple outcomes. For the $i$-th participant, let $p\in\{1,2,\ldots,r_i\}$ index the elements of the outcome. For example, in CHEARS AAA, each participant is tested for both ears; that is, $r_i=2$ for each participant. Now the linear model is 
	\begin{align*}
		Y_{ip} = \sum_{j=1}^{M}\beta_jT_i^{(j)} + \gammavec^{\T} \Xvec_i +\etavec^\T\Uvec_{i,p}+ \epsilon_{ip},
	\end{align*}
	where $\Uvec_{i,p}$ is a vector of ear-level covariates. Let $\Yvec_{i}=(Y_{11},\ldots,Y_{ir_i})^\T$. The set of parameters $\thetavec=(\gammavec^\T,\betavec^\T,\etavec^\T)^T$ can be estimated by solving the generalized estimating equation (GEE) \shortcite{liang1986longitudinal,zeger1986longitudinal}
	\begin{align}
		\sum_{i=1}^{N}\Dvec^\T_{i}(\thetavec)\vmat^{-1}_i(\thetavec,\nuvec)[\Yvec_{i}-\muvec_i(\thetavec)]=0,\label{eq:gee}
	\end{align}
	where $\muvec_i(\thetavec)=\E[\Yvec_{i}|\Xvec_i,\Uvec_{i,p},T_i^{(1)},\ldots,T_i^{(M)}]$, $\Dvec_i=\frac{\partial}{\partial\thetavec}\muvec_i(\thetavec)$, $\vmat_i(\thetavec,\nuvec)$ is the working variance-covariance matrix of the vector outcome for the $i$-th participant, and $\nuvec$ is the set of parameters for the correlation among multiple outcome measurements. Some commonly-used correlation structures are independent ($\text{Cor}(Y_{ip_1},Y_{ip_2})=0$ for $p_1\neq p_2\in\{1,\ldots,r_i\}$), exchangeable ($\text{Cor}(Y_{ip_1},Y_{ip_2})=\nu$), and unstructured ($\text{Cor}(Y_{ip_1},Y_{ip_2})=\nu_{p_1,p_2}$). The variance-covariance matrix of $\wh\betavec$, ${\text{Var}} ({\wh\betavec})$, can be estimated based on the sandwich variance approach \shortcite{liang1986longitudinal,zeger1986longitudinal} or the model-based variance. The former is robust to misspecification of the working variance-covariance matrix and the latter assumes that the working variance-covariance matrix is correctly specified. In this paper, we present the results based on both the sandwich and model-based variance estimates of $\text{Var}(\hat\betavec)$.
	
	After obtaining the evaluators' effects, $\hat{\betavec}$, the outlier evaluators are those with different evaluator effects. It is common that in a large epidemiological study, there is more than one outlier evaluator. In the next section, we propose a procedure that can detect the outliers in sequence based on $\hat{\betavec}$ and ${\text{Var}} ({\wh\betavec})$.
	
	\subsection{Stage II: Stepwise Hypothesis Testing Procedure}
	Based on the asymptotic property of the GEE-estimators (\shortciteNP{liang1986longitudinal,zeger1986longitudinal}), when $N$ is large, $\wh\betavec$ follows a multivariate normal distribution,
	\begin{align*}
		\wh\betavec\sim\NOR\left(\betavec,\omegamat_{\betavec}\right),
	\end{align*}
	where $\omegamat_{\betavec}=\text{Var}(\hat\betavec)$. The $j$-th evaluator is treated as an outlier when the corresponding effect $\beta_j$ is different from the effects of other normal evaluators.
	
	\citeN{rosner1983percentage} provides an outlier-detection procedure, which, however, can only be applied to independent observations. We extend that method to detect the outliers among $\{\hat\beta_1,\hat\beta_2,\ldots,\hat\beta_M\}$, which are correlated. We first specify the maximum number of potential outliers, denoted as $k$. The proposed outlier detection procedure has $k$ steps, and one potential outlier is tested at each step. At the initial step (i.e. $t=1$), the candidate set where the potential outliers will be selected from is $\Iset_1=\{1,\ldots,M\}$, and the index of the potential outlier that is identified from $\Iset_1$ is denoted as $o_1$. At the $t$-th step, $t=2,\ldots,k$, the index of the $t$-th potential outlier identified from $\Iset_{t}$ is denoted as $o_t$. To identify the potential outlier in $\Iset_{t}$, we use the idea of the modified Extreme Studentized Deviate (mESD) \cite{crosby1994detect}:
	\begin{align*}
		R_t = \max_{m\in \Iset_t}\frac{\left[\hat{\beta}_m-\frac{1}{|\Iset_{t}|}\sum_{h\in\Iset_{t}}\hat{\beta}_h\right]^2}{\text{var}\left[\hat{\beta}_m-\frac{1}{|\Iset_{t}|}\sum_{h\in\Iset_{t}}\hat{\beta}_h\right]},
	\end{align*}
	where $|\Iset_{t}|$ is the number of elements in set $\Iset_{t}$, $\frac{1}{|\Iset_{t}|}\sum_{j\in\Iset_{t}}\hat{\beta}_j$ is the mean of $\wh\betavec_{\Iset_t}$, and $\wh\betavec_{\Iset_t}$ is the vector of the elements in $\wh\betavec$ with indexes in $\Iset_{t}$. To reduce the possibility that the mean is largely impacted by a few extreme values in $\hat{\beta}_{\Iset_{t}}$, we can also replace $\frac{1}{|\Iset_{t}|}\sum_{j\in\Iset_{t}}\hat{\beta}_j$ with the truncated mean of $\betavec_{\Iset_t}$, denoted as $\text{trunc}(\wh\betavec_{\Iset_t})$. Let $\hat\beta_{(1)},\hat\beta_{(2)},\ldots,\hat\beta_{(|\Iset_t|)}$ be the ascendingly ordered values of $\wh\betavec_{\Iset_t}$. The $\delta\times100\%$ truncated mean is
	\begin{align*}
		\text{trunc}(\wh\betavec_{\Iset_t})=\frac{1}{|\Iset_t|-2\left[\delta\cdot|\Iset_t|\right]}\sum_{h=\left[\delta\cdot|\Iset_t|\right]+1}^{|\Iset_{t}|-\left[\delta\cdot|\Iset_t|\right]}\hat\beta_{(h)},
	\end{align*}
	where $\left[\delta\cdot|\Iset_t|\right]$ is the integer part of $\delta\cdot|\Iset_t|$. Using this truncated mean, mESD becomes 
	\begin{align*}
		R_t= \max_{m\in \Iset_t}\frac{\left[\hat{\beta}_m-\text{trunc}(\wh \betavec_{\Iset_t})\right]^2}{\text{var}\left[\hat{\beta}_m-\text{trunc}(\wh \betavec_{\Iset_t})\right]}.
	\end{align*}
	If $k$ is much smaller than $M$, which is typically the case in practice, we may choose to exclude the largest and smallest $k$ evaluators from calculating the mean so that the outlier evaluators' effects are less likely to have an impact on the truncated mean. 
	
	We define the contrast vector $\Lvec_{m,t}\in\mathbb{R}^{M-t+1}$ such that $\Lvec_{m,t}^\T\hat\betavec_{\Iset_t}=\hat{\beta}_m-\frac{1}{|\Iset_{t}|}\sum_{h\in\Iset_{t}}\hat{\beta}_h$. It follows that the $(M-t+1)$-element vector $\Lvec_{m,t}$ has the $m$-th element equal to $1-1/|\Iset_{t}|$ and all the other elements equal to $-1/|\Iset_{t}|$. Let the set of regression coefficients that are truncated be $\mathcal{A}_t=\{\beta_{(1)},\ldots,\beta_{(\left[\delta\cdot|\Iset_t|\right])},\beta_{(M-\left[\delta\cdot|\Iset_t|\right]+1)},\ldots,\beta_{(M)}\}$, then the $h$-th element of $\Lvec_{m,t}$ when using the truncated mean is
	\begin{align*}
		\left(\Lvec_{m,t}\right)_{h}=
		\begin{cases}
			0,& \text{if } \beta_h\in\mathcal{A}_t,\text{ and }h\neq m\\
			1,              & \text{if } \beta_h\in\mathcal{A}_t,\text{ and }h= m\\
			-\frac{1}{M-2\left[\delta\cdot|\Iset_t|\right]},& \text{if } \beta_h\notin\mathcal{A}_t,\text{ and }h\neq m\\
			1-\frac{1}{M-2\left[\delta\cdot|\Iset_t|\right]},              & \text{if } \beta_h\notin\mathcal{A}_t,\text{ and }h= m.
		\end{cases}
	\end{align*}
	In this paper, we use the truncated mean to calculate mESD. Then we can rewrite $R_t$ as
	\begin{align*}
		R_t=\max_{m\in \Iset_t}\frac{\left(\Lvec_{m,t}^\T\wh \betavec_{\Iset_t}\right)^2}{\Lvec_{m,t}^\T\omegamat_{\betavec_{\Iset_t}}\Lvec_{m,t}},
	\end{align*}
	where $\omegamat_{\betavec_{\Iset_t}}$ is variance-covariance matrix of $\wh\betavec_{\Iset_t}$. The index of the $t$-th potential outlier $\beta^{o_t}$ selected at the $t$-th step is
	\begin{align*}
		{o_t}=\arg\max_{m\in \Iset_t}\frac{\left(\Lvec_{m,t}^\T\wh \betavec_{\Iset_t}\right)^2}{\Lvec_{m,t}^\T\omegamat_{\betavec_{\Iset_t}}\Lvec_{m,t}}.
	\end{align*}
	
	We propose a stepwise Hypothesis Testing procedure. The null and alternative hypotheses at the $t$-th step are
	\begin{align*}
		H_{0,t}:|\Lvec_{m,t}\wh{\betavec}_{\Iset_t}|=0,\text{ for any } m\in\Iset_{t},\;\text{v.s.}\;H_{a,t}:\text{there is at least one }m\in\Iset_{t}\;\text{s.t.}\;|\Lvec_{m,t}\wh{\betavec}_{\Iset_t}|=c>0.
	\end{align*}
	If $R_{o_t}$ is greater than the critical value, the null hypothesis $H_{0,t}$ will be rejected and the $o_t$-th evaluators will be detected as an outlier. 
	
	Under the type I error rate $\alpha$, the critical value, $\lambda_t$, at the $t$-th step, $t=1,\ldots,k$, can be determined by the following equation,
	\begin{align*}
		\pr\left[\bigcup_{l=t}^{k}(R_l>\lambda_l)\given[\Big] \Hset_{t-1}\right] = \alpha,\,t=1,\ldots,k,
	\end{align*}
	where $\Hset_0$ is the null hypothesis of no outliers, and $\Hset_{t-1}$ for $t=2,\ldots,k$ represents the hypothesis that there are exactly $t-1$ outliers among the evaluators and all those outliers have been removed in previous $t-1$ steps. We use the approximation formula in \citeN{rosner1983percentage} to derive the critical value:
	\begin{align}
		1-\alpha=\pr\left[\bigcap_{l=t}^{k}(R_l\leq\lambda_l)\given[\Big] \Hset_{t-1}\right]\approx\pr\left(R_{t}\leq\lambda_{t}\given[\Big] \Hset_{t-1}\right) ,\,t=1,\ldots,k. \label{eq:alp}
	\end{align}
	It follows that $\lambda_t$ is the $1-\alpha$ quantile of $R_t$. A detailed derivation for the quantile of $R_t$ is provided in the appendix. Given the critical values $\lambda_1,\ldots,\lambda_k$, the number of detected outliers, $k'$, is the largest index among those with the outlier statistic, $R_t$, greater than the corresponding critical value $\lambda_{t}$, $t=1,\ldots,k$; that is,
	\begin{align*}
		k'=\max\,l,\text{ s.t.}\, R_l>\lambda_l.
	\end{align*}
	The $k'$ detected outlier evaluators are $\beta^{o_1},\ldots,\beta^{o_{(k')}}$. 
	
	\section{Simulation Study}
	We conducted a simulation study to evaluate the finite sample performance of our outlier detection method. The simulation settings were similar to CHEARS AAA data, including both single outcome and multiple outcomes scenarios. We evaluated the performance of our method under two scenarios, with and without outlier evaluators. For the scenario without outliers, we evaluated the type I error rates. For the scenario with outliers, we considered the true positive rate (TPR) and true negative rate (TNR) defined below:
	\begin{align*}
		\text{TPR}=&\frac{\text{the number of outliers detected as outliers}}{\text{the number of outliers}},\\
		\text{TNR}=&\frac{\text{the number of normal evaluators detected as normal evaluators}}{\text{the number of normal evaluators}}.
	\end{align*}
	
	\subsection{Single Measurement}
	In this setting, each participant had one measurement. We generated data mimicking the CHEARS AAA dataset. The linear model for generating the hearing measurements was:
	\begin{align*}
		Y_i=\eta_1\text{age}_i+\eta_2\text{age}_i^2+\eta_3\delta(\text{very good}_i)+\eta_4\delta(\text{a little hearing trouble}_i)+\sum_{j=1}^M\beta_jT_i^{(j)}+\epsilon_i.
	\end{align*}
	Each of 50 evaluators ($M=50$) tested 120 participants, and 
	$\delta(\text{very good}_i)$ and\\ $\delta(\text{a little hearing trouble}_i)$ were the indicators of the $i$-th participant's self-reported hearing status, with ``excellent'' being the reference group. The values of coefficients are $(\eta_1,\eta_2,\eta_3,\eta_4)=(-2.73,0.03,0.03,3.32)$, following the corresponding estimates in CHEARS AAA.
	
	To generate $Y_i$, the random noise $\epsilon_i, i=1,\ldots,N,$ followed an independent and identical normal distribution with zero mean and standard deviation $\sigma$ varying from 2 to 10. We set the age of each participant following a normal distribution with a mean of 56.56 years and a standard deviation of 4.36 years to mimic the CHEARS AAA, and the prevalences of the categories ``very good'' and ``a little hearing trouble'' were 0.44 and 0.25.
	
	\subsubsection{Type I Error Rate}
	We first studied the type I error rate. In this simulation study, the evaluator effect, $\beta_j,$ was set to 66.95 for all $j$'s. We set the maximum number of potential outliers $k=10$. Shown in Table~\ref{tab:singlealp} is the type I error rate (i.e. proportion of the simulation replicates with at least one outliers detected) for significance levels, $\alpha$, from 0.05 to 0.3, under 5000 simulation replicates. The type I error rates were all under significance level $\alpha$. As shown in Table~\ref{tab:singlealp}, the probability of misclassifying a normal evaluator as an outlier did not necessarily increase when the variance of the random noise increased.
	
	\subsubsection{True Positive Rate and True Negative Rate}
	In this simulation study, we evaluated the TNR and TPR of our outlier-detection procedure. Among the 50 evaluators, 10 had a different outlier effect; 5 of the 10 outliers' effects were significantly different from the normal evaluators, and the other 5 were intermediate outliers. The values of the evaluators' effects for the normal evaluator, intermediate outlier, and significant outlier were 66.95, 70.10, and 75.10 respectively.
	
	The significance levels ranged from 0.05 to 0.3, and $\sigma$ from 2 to 10. The maximum number of potential outliers was set to $k=10$. Shown in Table~\ref{tab:singlepow} are the TNR and TPR. Our proposed procedure had a satisfactory TNR in all the simulation settings, which suggested that the proposed outlier-detection procedure rarely misclassified a normal point as an outlier. For TPR, when the random noise's standard deviation $\sigma$ was $10$, the proposed outlier-detection procedure could typically detect the 5 significant outliers, while when $\sigma$ was less than 6, all remaining 5 intermediate outliers could be detected. The simulation results for increased values of $\sigma$ (2 to 14), $\alpha$ (0.05 to 0.3), and $k$ (10 and 15) in Table 1-2 of the supplemental material were similar.
	
	\subsection{Multiple Outcome Measurements}\label{sec:multi}
	In CHEARS AAA, each participant was tested for both ears. We designed the simulation studies with multiple outcomes similar to the CHEARS AAA data. The outcomes from the left and right ear of each participant were correlated and followed a normal distribution. The linear model for generating the clustered hearing measurements was
	\begin{align*}
		&Y_{ip}=\eta_1\text{age}_i+\eta_2\text{age}_i^2+\eta_3\delta(\text{very good}_i)+\eta_4\delta(\text{a little hearing trouble}_i)+\sum_{j=1}^M\beta_jT_i^{(j)}+\epsilon_{ip}, p=1,2,\\
		&\begin{pmatrix}\epsilon_{i1}\\
			\epsilon_{i2}
		\end{pmatrix} \sim \NOR
		\begin{bmatrix}
			\begin{pmatrix}
				0\\
				0
			\end{pmatrix}\!\!,&\sigma^2
			\begin{pmatrix}
				1 & \rho \\
				\rho& 1 
			\end{pmatrix}
		\end{bmatrix}.\\[2\jot]
	\end{align*}
	The values of $\eta_1,\ldots,\eta_4$ were the same as in Section 3.1. We set the correlation coefficient $\rho$ to be 0.3, 0.5, and 0.8, representing the low, moderate, and strong correlation between ears, and used the exchangeable working variance-covariance matrix in the GEE analysis. Same as the single outcome measurement scenario, we had 50 evaluators, and each evaluator tested 120 participants.
	
	\subsubsection{Type I Error Rate}
	To study the type I error rate for the multiple measurements scenario, we set the maximum number of potential outliers $k=10$. Table~\ref{tab:multialp} shows the type I error rate under the significance levels from 0.05 to 0.3 using the model-based and sandwich variance estimation of $\text{Var}(\hat{\betavec})$, both under 5000 simulation replicates. The critical values using both the model-based and sandwich variance estimation had the type I error rates close to the significance level.
	
	\subsubsection{True Positive Rate and True Negative Rate}
	To study the TPR and TNR of the multiple measurements scenario, the values of evaluators' effects and numbers of the three types of evaluators were the same as in Section 3.1.2. The number of potential outliers $k$ was set to 10. Shown in Table~\ref{tab:multipow} are the TPR and TNR based on the model-based and sandwich variance estimation. The proposed method had satisfactory TNR and TPR, and TPR decreased when the noise $\sigma$ increased. In addition, since a lower correlation between two ears leads to more precise estimates in the first stage analysis, it yields a higher TPR for detecting outliers. The simulation results for increased values of $\sigma$ (2 to 14), $\alpha$ (0.05 to 0.3), and $k$ (10 and 15) in Table 3-10 of the supplemental material were similar.
	
	\section{Illustrative example}
	To illustrate our method, we applied the proposed outlier-detection procedure to detect potential audiologist outliers based on the hearing threshold measurements under 8kHz in CHEARS AAA measured in 2014. A total of 6398 participants were tested for both ears by 46 evaluators. In the first stage analysis, we used the GEE approach to estimate the coefficients of model~(\ref{eq:gee}). The maximum number of potential outliers was set to 17 and we used the sandwich variance estimation for $\text{Var}(\wh\betavec)$.
	
\begin{figure}
		\centering
		\includegraphics[width=0.6\linewidth]{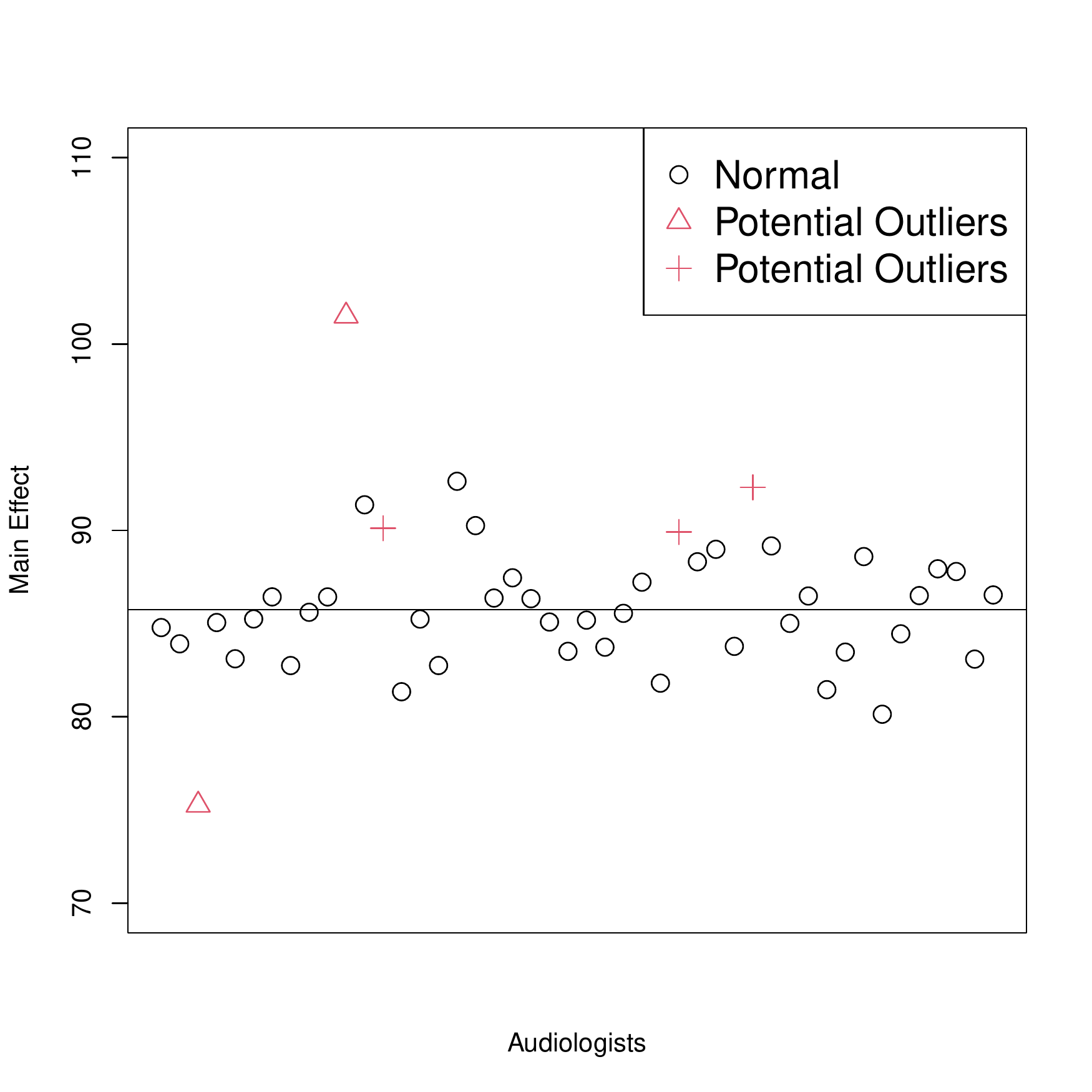}
		\caption{Estimated evaluators' effects $\hat{\betavec}$. The horizon line is the truncated mean of the effects. Potential outliers are in red. Two potential outliers detected under the type I error rate $\alpha=0.05$ are marked as red triangles. Three potential outliers detected under $\alpha$=0.4 are marked as red plus signs.}\label{fig:coef}
	\end{figure}

	Shown in Table~\ref{tab:outaudio} are the detected outlier evaluators under type I error rates, $\alpha$, ranging from 0.05 to 0.45. Two potential outliers, 4 and 13, were detected when $\alpha=0.05$. When $\alpha$ increased from 0.05 to 0.40, 3 additional outliers, 41, 16, and 40, were selected. Figure~\ref{fig:coef} shows the estimated effects of evaluators, with the red triangle markers representing the potential outliers, 4 and 13, selected under $\alpha =0.05$, and red plus markers for the additional potential outliers, 41, 16, 40, selected under $\alpha=0.40$. The device used by Audiologist 13 was found flawed in a quality control examination triggered by the fact that this audiologist was classified as a potential outlier.
	%We can see most evaluators with effects significantly far from the truncated mean are detected as outliers. Only 2 evaluators are significant and the rest 7 have similar effects compared to the rest of the normal evaluators. The 2 significant outliers in Figure~\ref{fig:coef} is consistent with the results in Table~\ref{tab:outaudio} as evaluators 4 and 13 can be detected when $\alpha$ equals to 0.05.
	
	\section{Discussion}
	In this paper, we provided a two-stage stepwise hypothesis testing approach for detecting potential outliers among evaluators. In the first stage, we used a regression model to estimate evaluators' effects. In the second stage, we proposed a stepwise hypothesis testing procedure to detect outliers among evaluators' estimated effects. We derived an approximation formula to calculate the critical values. The finite sample performance of this approach was evaluated in a simulation study, where the type I error rate, TPR, and TNR were satisfactory. An increase in the noise variance decreased TPR; however, the type I error rate remained the same. 
	
	In the simulation study for the multiple outcomes scenario, the model-based variance and sandwich variance had a similar performance. The model-based variance is only valid when the working variance-covariance matrix $\vmat(\thetavec,\nuvec)$ in Equation~(\ref{eq:gee}) is correctly specified while the sandwich variance is valid as long as the mean model is correct. Therefore, the sandwich variance estimation is preferable for the multiple outcome scenarios unless the structure of $\vmat(\thetavec,\nuvec)$ is known. In our simulation study, a larger maximum number of potential outliers, $k$, had a similar performance to the situation where $k$ was set to 10, the exact number of outliers. Specifically, TNR did not decrease when $k$ was larger than the number of outliers.
	
	The R code for the proposed approach is available at \url{https://www.hsph.harvard.edu/molin-wang/software/}.

	\section*{Acknowledgment}
	This work is supported by NIH grant R01DC017717.
	
	\clearpage
	%%%%%%%%%%%%%%%%%%%%%%%%%%%%%%%%%%%%%%%%%%%%%%%%%%%%%%%%%%%%%%%%%%%%%%%%%%%%%%%%%%%%%%%%%%%%%%%%%%%%%%%%%%%%%%%%%%%%%
	\bibliographystyle{chicago}
	\bibliography{ref}
	\clearpage
	\begin{table}
		\centering
		\caption{Type I error rate for the single measurement simulation.} \label{tab:singlealp}
		\begin{tabular}{llll}
			\hline
			$\sigma$    & 2        & 6        & 10         \\\hline
			$\alpha=$0.05 & 0.051 & 0.039 & 0.041 \\ 
			$\alpha=$0.1 & 0.087 & 0.084 & 0.089 \\ 
			$\alpha=$0.3 & 0.279 & 0.268 & 0.266 \\ \hline
			\multicolumn{4}{l}{$\sigma$ is the standard deviation of}\\
			\multicolumn{4}{l}{random noise;}\\
			\multicolumn{4}{l}{$\alpha$ is the significance level.}
		\end{tabular}
	\end{table}
	% \text{TPR}=\frac{\#(\text{Outlier detected as Outlier})}{\#(\text{Outliers})},\\
	% \text{TNR}=\frac{\#(\text{Normal detected as Normal})}{\#(\text{Normal})}.
	
	\begin{table}
		\centering
		\caption{True negative rate (TNR) and True positive rate (TPR) for the single measurement simulation; 10 outliers among 50 evaluators.}\label{tab:singlepow}
		\begin{tabular}{lllllll}%
			\hline
			$\sigma$ & 2  & 6  & 10 & 2  & 6  & 10  \\\hline
			&\multicolumn{3}{c}{TPR}&\multicolumn{3}{c}{TNR}\\\hline
			$\alpha$=0.05 & 1.000 & 0.996 & 0.781 & 1.000 & 1.000 & 0.999\\ 
			$\alpha$=0.1 & 1.000 & 0.997 & 0.822& 1.000 & 1.000 & 0.998 \\ 
			$\alpha$=0.3 & 1.000 & 0.998 & 0.879  & 1.000 & 1.000 & 0.994\\ 
			\hline
			\multicolumn{7}{l}{$\text{TNR}=\frac{\text{the number of normal evaluators detected as normal evaluators}}{\text{the number of normal evaluators}}$;}\\
			\multicolumn{7}{l}{$\text{TPR}=\frac{\text{the number of outliers detected as outliers}}{\text{the number of outliers}}$;}\\
			\multicolumn{7}{l}{$\alpha$ is the significance level;}\\
			\multicolumn{7}{l}{$\sigma$ is the standard deviation of random noise.}
		\end{tabular}
	\end{table}
	
	% \usepackage{multirow}
	%	\begin{table}
		%		\centering
		%		\caption{Mean number of evaluators detected as outliers among 10 outliers for the single measurement simulation.}\label{tab:nrej}
		%		\begin{tabular}{lllllllll}
			%			\hline
			%			\multicolumn{2}{c}{$\sigma^2$}      & 2      & 4      & 6     & 8     & 10    & 12    & 14    \\\hline
			%			\multirow{10}{*}{$\alpha$} & 0.05 & 10.055 & 9.675  & 7.123 & 5.758 & 5.044 & 3.899 & 2.755 \\
			%			& 0.1  & 10.091 & 9.86   & 7.498 & 6.076 & 5.34  & 4.431 & 3.204 \\
			%			& 0.15 & 10.148 & 9.983  & 7.931 & 6.285 & 5.568 & 4.671 & 3.681 \\
			%			& 0.2  & 10.23  & 10.075 & 8.12  & 6.525 & 5.721 & 4.914 & 3.923 \\
			%			& 0.25 & 10.267 & 10.131 & 8.322 & 6.669 & 5.873 & 5.099 & 4.141 \\
			%			& 0.3  & 10.332 & 10.212 & 8.607 & 6.934 & 6.083 & 5.326 & 4.318 \\
			%			& 0.35 & 10.439 & 10.314 & 8.712 & 7.083 & 6.168 & 5.482 & 4.544 \\
			%			& 0.4  & 10.486 & 10.421 & 8.813 & 7.272 & 6.377 & 5.647 & 4.738 \\
			%			& 0.45 & 10.559 & 10.466 & 9.05  & 7.493 & 6.524 & 5.815 & 4.878 \\
			%			& 0.5  & 10.642 & 10.614 & 9.337 & 7.575 & 6.77  & 5.923 & 5.143 \\\hline
			%		\end{tabular}
		%	\end{table}
	
	\begin{table}[]
		\centering
		\caption{Type I error rate for the multiple measurements simulation using the model-based and sandwich variance estimation in the first stage GEE analysis; significance level $\alpha$=0.05.} \label{tab:multialp}
		\begin{tabular}{lllllll}
			\hline
			$\sigma$& 2 & 6 & 10 & 2 & 6 & 10 \\ 
			\hline
			&\multicolumn{3}{c}{Model-based Variance}&\multicolumn{3}{c}{Sandwich Variance}\\
			$\rho=$0.3 & 0.042 & 0.043 & 0.043& 0.058 & 0.061 & 0.057 \\ 
			$\rho=$0.5 & 0.043 & 0.040 & 0.044& 0.059 & 0.060 & 0.056 \\ 
			$\rho=$0.8 & 0.042 & 0.048 & 0.049& 0.059 & 0.059 & 0.060 \\ 
			\hline
			\multicolumn{7}{l}{$\sigma$ is the standard deviation of random noise;}\\
			\multicolumn{7}{l}{$\rho$ is the correlation between right and left ears.}
		\end{tabular}
	\end{table}
	
	\begin{table}[ht]
		\centering
		\caption{True negative rate (TNR), and true positive rate (TPR) for the multiple measurements simulation using the model-based and sandwich variance estimation in the first stage of GEE analysis; 10 outliers among 50 evaluators; significance level $\alpha$=0.05.}\label{tab:multipow}
		\begin{tabular}{lllllll}%{llll}
			\hline
			$\sigma$& 2  & 6 & 10& 2  & 6 & 10 \\ 
			\hline
			&\multicolumn{3}{c}{TPR}&\multicolumn{3}{c}{TNR}\\
			\multicolumn{7}{c}{Model-based Variance}\\
			$\rho=$0.3 & 1.000 & 1.000 & 0.908 & 1.000 & 1.000 & 1.000 \\ 
			$\rho=$0.5 & 1.000 & 1.000 & 0.859 & 1.000 & 1.000 & 1.000 \\ 
			$\rho=$0.8 & 1.000 & 0.998 & 0.797 & 1.000 & 1.000 & 1.000 \\ 
			\hline
			\multicolumn{7}{c}{Sandwich Variance}\\
			$\rho=$0.3 & 1.000 & 1.000 & 0.906 & 1.000 & 1.000 & 1.000\\ 
			$\rho=$0.5 & 1.000 & 1.000 & 0.865 & 1.000 & 1.000 & 0.999\\ 
			$\rho=$0.8 & 1.000 & 0.998 & 0.798 & 1.000 & 1.000 & 0.999 \\ 
			\hline
			%			\multicolumn{8}{c}{Average detection}\\\hline
			%			$\rho=0.3$ & 10.050 & 10.026 & 8.700 & 6.713 & 5.763 & 5.180 & 4.437 \\ 
			%			$\rho=0.5$ & 10.057 & 9.990 & 8.162 & 6.294 & 5.546 & 4.903 & 3.935 \\ 
			%			$\rho=0.8$ & 10.049 & 9.856 & 7.480 & 5.963 & 5.261 & 4.386 & 3.223 \\ 
			%			\hline
			\multicolumn{7}{l}{$\text{TNR}=\frac{\text{the number of normal evaluators detected as normal evaluators}}{\text{the number of normal evaluators}}$;}\\
			\multicolumn{7}{l}{$\text{TPR}=\frac{\text{the number of outliers detected as outliers}}{\text{the number of outliers}}$;}\\
			\multicolumn{7}{l}{$\sigma$ is the standard deviation of random noise;}\\
			\multicolumn{7}{l}{$\rho$ is the correlation between right and left ears.}
		\end{tabular}
	\end{table}
	
	\begin{table}[]
		\centering
		\caption{Detected outlier evaluators under various type I error rates, $\alpha$, for the CHEARS AAA hearing measurements in 2014.}\label{tab:outaudio}
		\begin{tabular}{ll}
			\hline
			$\alpha$ & Selected Audiogolists               \\\hline
			0.05     & 4, 13                             \\
			0.10     & 4, 13, 41                         \\
			%0.15     & 4, 13, 41, 16                     \\
			0.20     & 4, 13, 41, 16                     \\
			%0.25     & 4, 13, 41, 16, 40                 \\
			0.30     & 4, 13, 41, 16, 40                 \\
			%0.35     & 4, 13, 41, 16, 40                 \\
			0.40     & 4, 13, 41, 16, 40                 \\
			%0.45      & 4, 13, 41, 16, 40, 17, 48, 22, 24 \\
			\hline
		\end{tabular}
	\end{table}
	
	\clearpage
	\appendix
	\section*{Technical Details on Deriving the Critical Values}\label{sec:app}
	Here, we show how to derive the critical values $\lambda_{t},$ $t=1,\ldots,k$. From Equation~(\ref{eq:alp}), we have
	\begin{align*}
		1-\alpha
		&=\pr\left[\bigcap_{l=t}^{k}(R_l\leq\lambda_l)\given[\Big] \Hset_{t-1}\right]\approx\pr\left(R_{t}\leq\lambda_{t}\given[\Big] \Hset_{t-1}\right)\\
		&=\Pr\left(\max_{m\in \Iset_{t}}\frac{\left(\Lvec_{m,t}^\T\wh\betavec_{\Iset_{t}}\right)^2}{\Lvec_{m,t}^\T\omegamat_{\betavec_{\Iset_{t}}}\Lvec_{m,t}}\leq\lambda_{t}\given[\Big] \Hset_{t-1}\right)\\
		&=\Pr\left[\bigcap_{m\in\Iset_{t}}\left(\frac{\left(\Lvec_{m,t}^\T\wh\betavec_{\Iset_{t}}\right)^2}{\Lvec_{m,t}^\T\omegamat_{\betavec_{\Iset_{t}}}\Lvec_{m,t}}\leq\lambda_{t}\right)\given[\Big] \Hset_{t-1}\right],\text{ for }t=1,\ldots,k.
	\end{align*}
	Since $\wh\betavec_{\Iset_{t}}$ follows the multivariate normal distribution when $N$ is large,
	\begin{align*}
		\wh \betavec_{\Iset_{t}}\sim\NOR\left(\betavec_{\Iset_{t}}, \omegamat_{\betavec_{\Iset_{t}}}\right).
	\end{align*}
	Define $\Zvec=(Z_1,\ldots,Z_{M-t+1})^\T$ such that $\Zvec=\Amat_t\wh\betavec_{\Iset_{t}}$ and $\Amat_t$ is a matrix with rows equals to $\frac{\Lvec_{m,t}^\T}{\sqrt{\Lvec_{m,t}^\T\omegamat_{\betavec_{\Iset_{t}}}\Lvec_{m,t}}}$ for $m\in \Iset_{t}$. Then 
	\begin{align*}
		\Zvec\given \Hset_{t-1}\sim\NOR\left(0,\Amat_t\omegamat_{\betavec_{\Iset_{t}}}\Amat_t^\T\right),\\
		R_t=\max_{m\in \Iset_{t}}\frac{\left(\Lvec_{m,t}^\T\wh\betavec_{\Iset_{t}}\right)^2}{\Lvec_{m,t}^\T\omegamat_{\betavec_{\Iset_{t}}}\Lvec_{m,t}}=\max_{b=1}^{M-t-1}Z_b^2.
	\end{align*}
	To determine $\lambda_t$,
	\begin{align*}
		1-\alpha\approx\pr\left(R_{t}\leq\lambda_{t}\given[\Big] \Hset_{t-1}\right)\\
		=\Pr\left(\max_{b=1}^{M-t-1}Z_b^2\leq\lambda_t\right)=\Pr\left(\bigcap_{b=1}^{M-t-1}|Z_b|\leq\sqrt{\lambda_t}\right),
	\end{align*}
	so, $\sqrt{\lambda_t}$ is the $1-\alpha$ two-sided quantile of the distribution $\NOR\left(0,\Amat_t\omegamat_{\betavec_{\Iset_{t}}}\Amat_t^\T\right)$.
\end{document}